\documentstyle[twoside,epsf,fleqn,espcrc2]{article}

\newcommand{\AmS}{{\protect\the\textfont2
  A\kern-.1667em\lower.5ex\hbox{M}\kern-.125emS}}

\newcommand{\sesam}{{\small\sf SESAM }}

\newcommand{\sesamc}{{\small\sf SESAM} collaboration }

\newcommand{\ewxynarrow}[2]{\setlength{\epsfxsize}{#2}\epsfbox[10 30 560 590]{#1}}

\newcommand{\eq}{eq.~}
\newcommand{\beq}{\begin{equation}}
\newcommand{\eeq}{\end{equation}}
\newcommand{\bea}{\begin{eqnarray}}
\newcommand{\eea}{\end{eqnarray}}

\newcommand{\oks}{{1 \over \kappa_{\rm sea}}}

\newcommand{\mq}{m}
\newcommand{\mqv}{m^{\rm valence}}
\newcommand{\mqs}{m^{\rm sea}}
\newcommand{\mql}{m^{\rm light}}

\newcommand{\mot}{m_{\sf ss}}
\newcommand{\mtf}{m_{\sf vv}}
\newcommand{\moth}{m_{\sf sv}}

\newcommand{\mpsot}{m_{\rm PS, {\sf ss}}^2}
\newcommand{\mpstf}{m_{\rm PS, {\sf vv}}^2}
\newcommand{\mpsoth}{m_{\rm PS, {\sf sv}}^2}

\newcommand{\mvot}{m_{\rm V, {\sf ss}}}

\title{Light and Strange Quark Masses with Dynamical Wilson Fermions.}

\author{Henning Hoeber (for the {\small\sf SESAM}
        collaboration\thanks{N.~Eicker, U.~Gl\"assner, S.~G\"usken,
        H.~H, P.~Lacock, Th.~Lippert, G.~Ritzenh\"ofer, K.~Schilling, 
G.~Siegert, A.Spitz, P.~Ueberholz, J.~Viehoff.})\address{Department of
        Physics, University of Wuppertal, Gaussstr. 20, 42119 Wuppertal, Germany; \\
        HLRZ c/o J\"ulich Research Centre and DESY, Hamburg, 52425
        J\"ulich, Germany.}} 
\begin{document}
\begin{abstract}
We determine the masses of the light and the strange quarks in the
$\overline{MS}$-scheme using our high-statistics lattice simulation of QCD with
dynamical Wilson fermions. For each of our three sea quarks we have
analyzed our data at five different values of the valence quark mass,
enabling us to parameterize our fit results in the ($\mqs$, $\mqv$)
plane. For the light quark mass we find $m^{\rm light}_{\overline{MS}}(2\, {\sf GeV}) =
2.7(2)\, {\sf MeV}$, which is lower than in quenched
simulations. Applying a new method, which we propose to extract the
strange quark mass in a sea of two dynamical light quarks, we obtain
$m^{\rm strange}_{\overline{MS}}(2\, {\sf GeV}) = 140(20)\, {\sf MeV}$.
\end{abstract}
\maketitle
\section{Introduction}
More than 25 years after the introduction of quarks as the
constituents of hadrons the absolute values of the light quark masses,
as given in the current Particle Data Book \cite{PRD}, are still among
the most poorly known fundamental parameters of the Standard
Model. Chiral perturbation theory ($\chi$PT) has been very successful
in fixing the ratios of light quark masses \cite{leutwyler} but the
overall scale is beyond its possibilities. QCD sum-rules
\cite{sumrules} have produced a wealth of predictions but in a large
range of values and systematic errors are hard to pin down. Given the
shortcomings and problems of these two methods, the lattice, as a
first-principles non-perturbative method, would appear to be the most
promising candidate for a reliable extraction of the absolute quark
mass values. In this talk I present results of the \sesamc for the
masses of the light and strange quarks obtained from the currently
largest and statistically most precise lattice QCD simulation with 2
flavours of dynamical Wilson fermions \cite{us}.
\section{Method and Status}
Calculating quark  masses on the lattice consists of three steps:
firstly, a hadron mass is matched to its experimental value to fix the
bare lattice quark mass under consideration. Typically, the
pseudoscalar meson mass, having the smallest statistical lattice-error
of all hadrons, is used. This step also invokes $\chi$PT, often only
to lowest order, to relate the pseudoscalar mass to the bare quark
mass. Secondly, the lattice scale is set in a similar manner by
matching some experimentally well known quantity such as the $\rho$
mass or the $1S-2S$ energy level splitting in the $\Upsilon$
spectrum. Finally, perturbation theory is used to convert the bare
quark mass to that of the $\overline{MS}$ scheme.
\par
Quite a large number of simulations have been carried out in this
manner in the quenched approximation of QCD. Internal fermion loops
are totally neglected in these simulations for want of sufficient
compute power. Recent compilations and interpretations of these
results are given in \cite{mackenzie1,gupta} and \cite{allton1}. The
authors of \cite{mackenzie1,gupta} have each 
attempted to produce quark mass values in the limit of
vanishing lattice spacing. Although the analysis differ greatly in
approach and method the overall result is very much the same: the
isospin symmetric light quark and the strange quark masses are both
predicted by {\it quenched lattice QCD} to be smaller than the sumrule
results \cite{allton1}. For {\it full QCD}, very few results have been
obtained so far but these would seem to indicate even smaller quark
masses; however, no reliable continuum extrapolation can be performed
so far. 
\section{\sesam analysis}
\sesam has recently completed the generation of gauge
configurations with two dynamical Wilson fermions at three different
values of the dynamical mass and for a lattice extent of $16^3 \times
32$. The bare coupling of $\beta = 5.6$ corresponds to a cutoff
$a^{-1} \simeq 2.33(6) {\sf GeV}$ (matching the $\rho$)\footnote{If,
  instead, we match the $1S-2S$   splitting in the $\Upsilon$, the
  scale turns out to be $\simeq   2.48(14){\sf GeV}$.}. 200 lattices
are analysed per sea-quark mass, giving $m_{\pi}/m_{\rho} =
0.841(6),\;0.755(8),\; \mbox{and } 0.69(2)$. Correlators are evaluated
for 5 values of the valence quark mass at each sea quark mass value.
\subsection{Light Quark Mass}
\begin{figure}[tbp]
\centerline{
\ewxynarrow{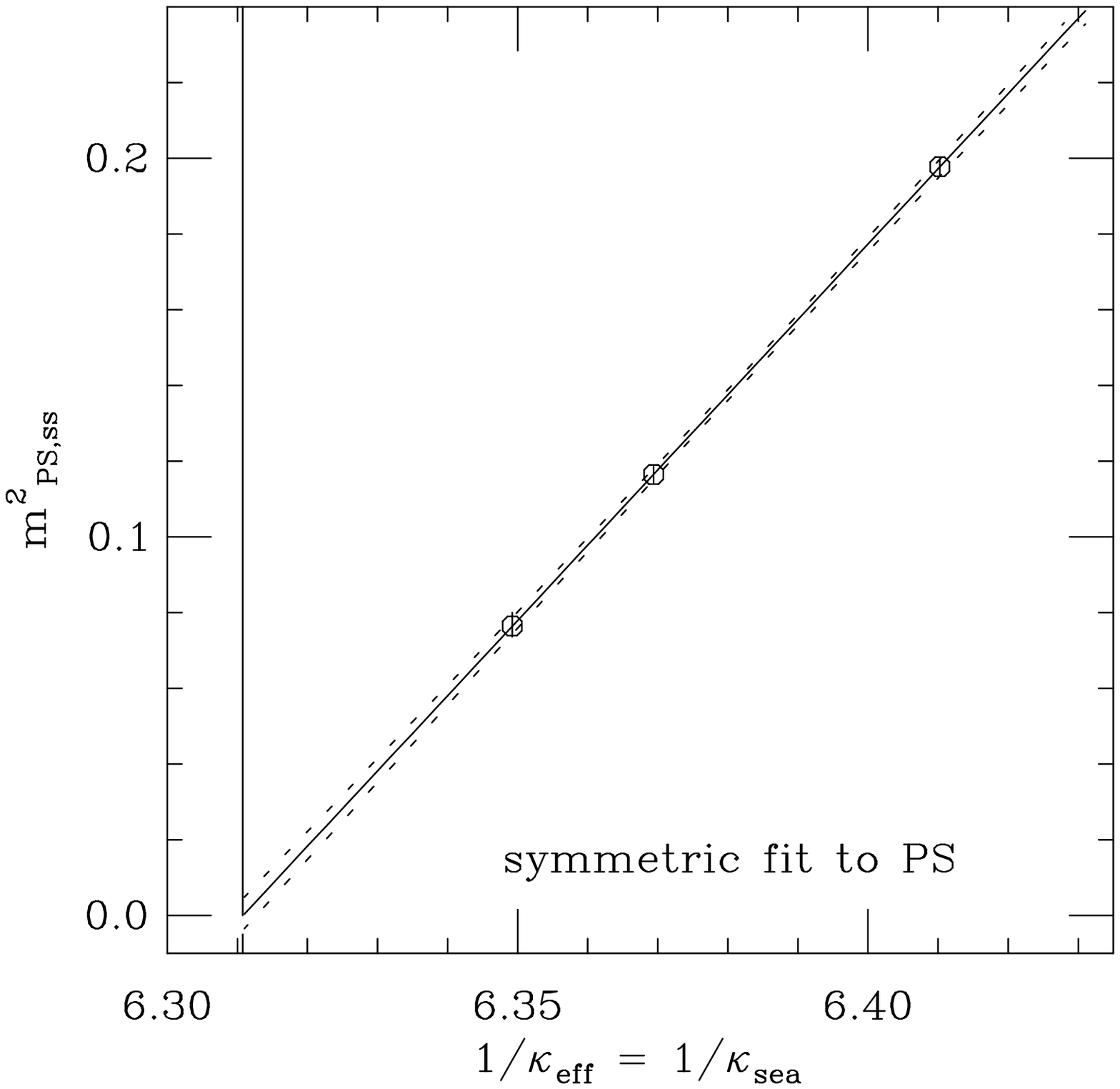}{250pt}       
}
\centerline{
\ewxynarrow{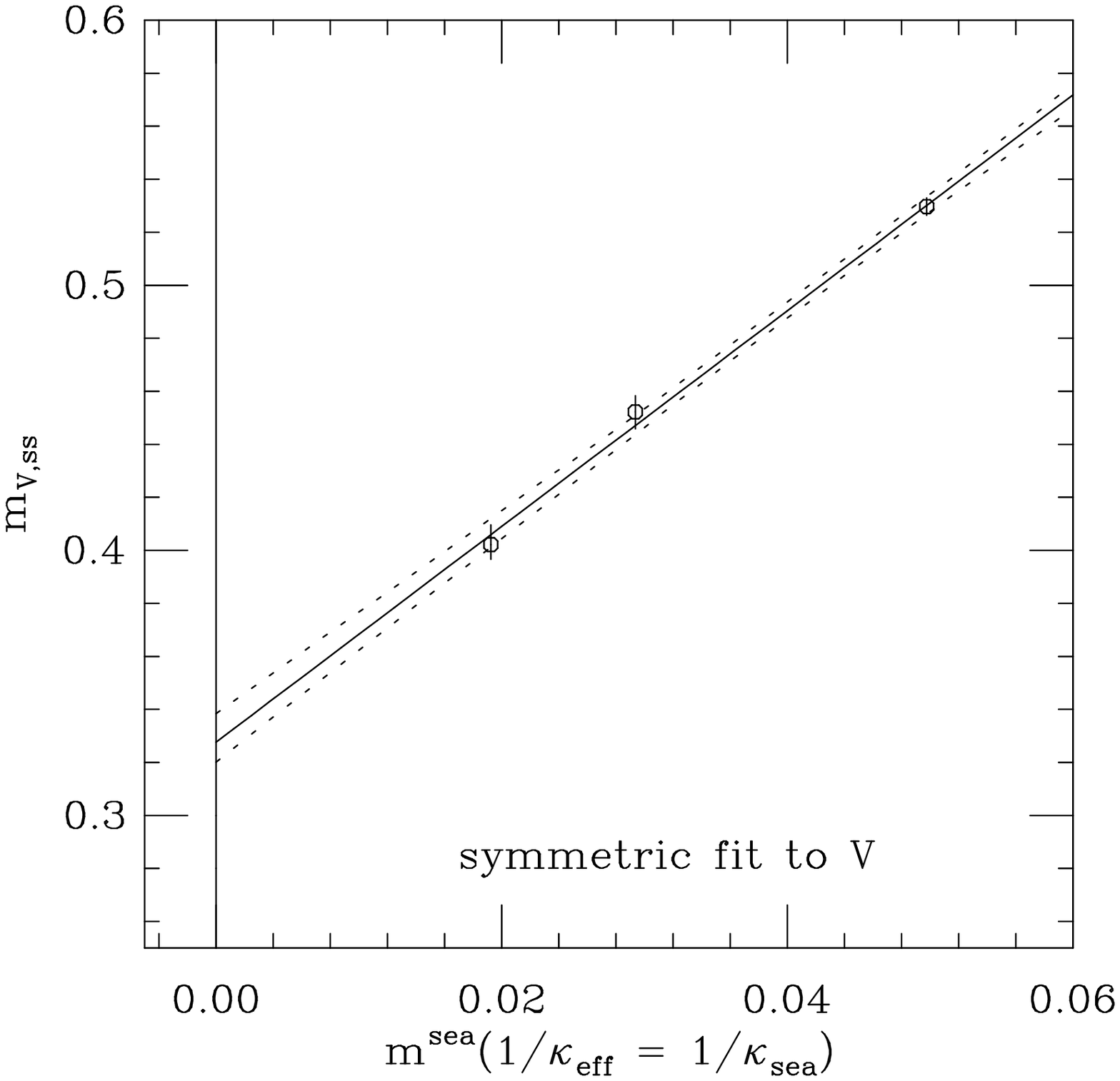}{250pt}
}
\caption{$\mpsot$ (left) as a function of $\oks$ and $\mvot$ as a
  function of $\mqs$ (in lattice units).
\label{fig1}}
\end{figure}

To extract the light quark mass from the pseudoscalar (PS) and vector (V)
meson mass ratio $m_\pi/m_\rho$ we attempt to fit the masses of mesons
with valence quark masses equal to sea quark masses (characterized by
the argument $ss$) according to first order $\chi$PT:
\begin{equation}
m_{PS}^2(ss) =  c\, m_q^{sea} \quad , \quad 
m_V(ss) = m^{crit} + b\, m^2_{PS} \label{eqn_symmetric} \; .
\end{equation}
Figure \ref{fig1} shows that this ansatz works very well (although a
quadratic term in the vector data cannot be excluded). The condition
\beq
\frac{m_{PS}(m_q^l)}{m_{V}(m_q^l)} = \left(\frac{m_{\pi}}{m_{\rho}}\right)
_{exp} = 0.1785 
\eeq
yields
\beq
m_{\overline{MS}}^{l}(2\, {\sf GeV}) = \frac{1}{2}(m_u + m_d) =
2.7(2) \, {\sf MeV}\; .
\label{eqn_light_cond}
\eeq
The lattice result has been converted to the ${\overline{MS}}$ scheme
using boosted 1-loop perturbation theory \cite{Zhang} and the scale was set by the
$\rho$-mass. 
%
\subsection{Strange Quark Mass}
So far, we have used our data with equal sea and valence quark masses to
extract the light quark mass in a sea of light quarks. To simulate the
strange quark we need to introduce valence quarks that are unequal to
the dynamical quarks. The
meson masses with strange valence quarks can then be extrapolated to
the sea of light quarks. To this end, encouraged by
eq.~\ref{eqn_symmetric}, we make the all-linear ansatz:
\bea
m_{PS}^2(sv) &=& c' m_q^{sea} + d' m_q^{val} \; , \label{sym1}\\
m_{PS}^2(vv) &=& c'' m_q^{sea} + \frac{1}{2}d''(m_q^{val1} + m_q^{val2}) 
\label{sym2}\;. 
\eea
The arguments $sv$ and $vv$ denote the valence quark content so that,
for example, $vv$ means neither of the valence quark masses is equal
to the sea quark mass. 
\par
Equations \ref{sym1} and \ref{sym2} seem to show that four additional
parameters are needed to parametrize the pseudoscalar meson mass in the
entire ($\mqs$, $\mqv$) plane. However, the consistency of equations
\ref{sym1} and {\ref{sym2} with eq.~\ref{eqn_symmetric} can be used to
  reduce the number of unknown parameters to 3 and similar arguments
  hold for the vector particle. Thus, we attempt fits to the combined
  parametrization 
\begin{eqnarray}
\left( 
\begin{array}{c} 
\mpsot  \\
\mpsoth \\
\mpstf(vv)  
\end{array}
\right) &=&  
\left( 
\begin{array}{cc} 
c & 0  \\
c  - d' & d' \\
c - d'' & d'' 
\end{array}
\right)   
\left( 
\begin{array}{c} 
\mqs \\
\mq^{\rm val.}
\end{array}
\right)   \;\; , 
\label{quarkmatrix}
\end{eqnarray}
and similarily for the vector meson. 
%
\begin{figure}[tbp]
\centerline{
\ewxynarrow{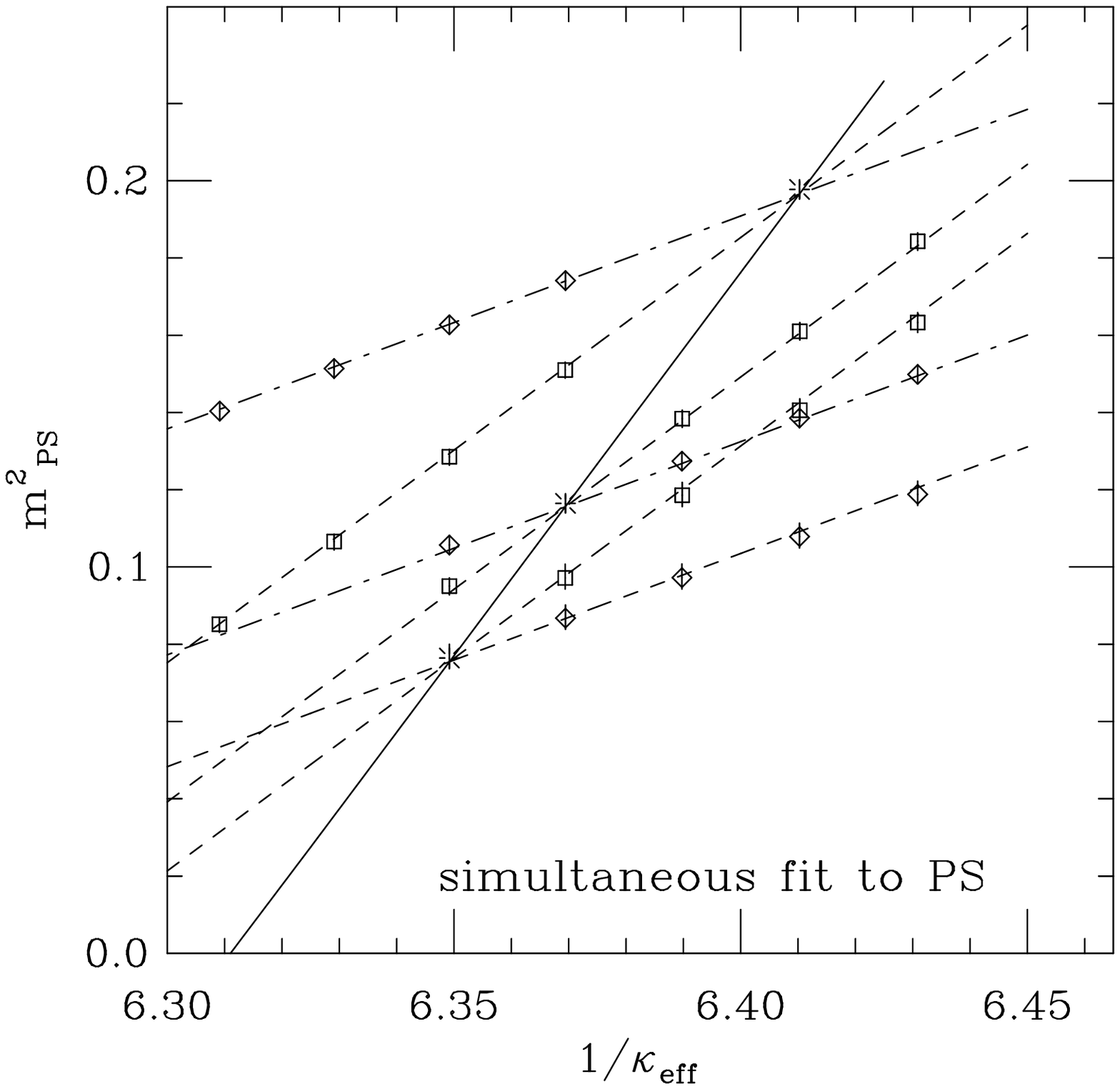}{250pt}
}
\centerline{
\ewxynarrow{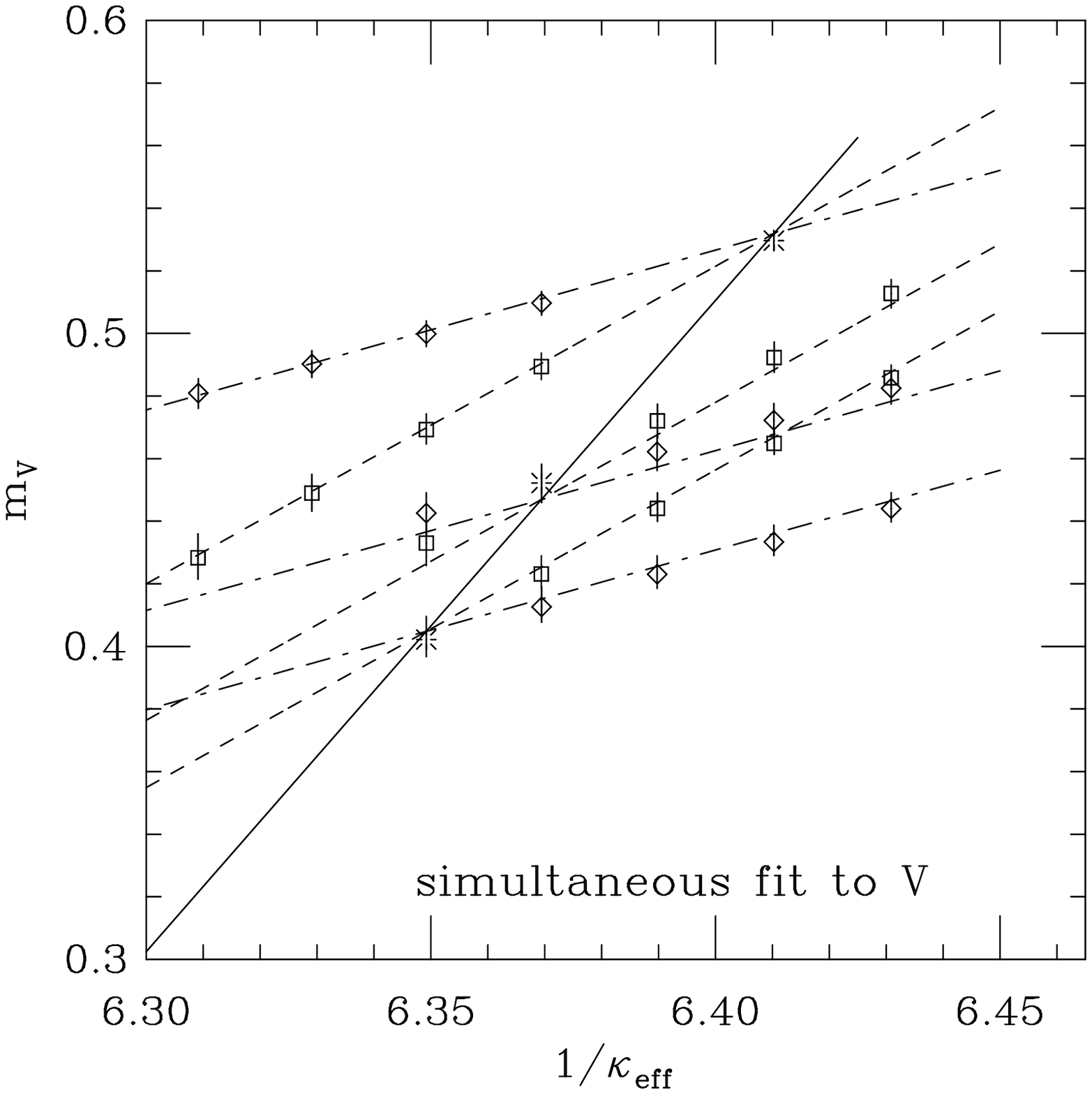}{250pt}       
}
\caption{Simultaneous fit of all pseudoscalar and vector data to
  \eq\ref{quarkmatrix}. Symbols: $* = \mot\; -\!\!-$; $\Diamond = \moth \; -
  \cdot\; -$; $\Box = \mtf \; - \, -$. All plots in lattice units.
\label{hugefit}}
\end{figure}
Figure \ref{hugefit} displays our data and the combined fits according
to this ansatz. Again, we find the parametrization to work very well
but cannot exclude small contributions from higher order $\chi$PT.
\par
The strange quark is now determined from the (redundant) system of
conditions
\bea
\left(\frac{m_{K^*}}{m_{\rho}}\right)_{exp}=1.16,  & 
\left(\frac{m_{K}}{m_{\rho}}\right)_{exp}=0.642\; ,\nonumber \\
\left(\frac{m_{\Phi}}{m_{\rho}}\right)_{exp}=1.326\;, & \nonumber
\eea
which has to be evaluated at $m_q^{sea} = m_q^l$. All three conditions
lead to consistent results for $m^s$; we average and include the
spread into the statistical error. After conversion into the
$\overline{MS}$ scheme we obtain 
\beq
m^s_{\overline{MS}}(2 \, {\sf GeV}) = 140(20)\, {\sf MeV}\;.
\label{strange_final}
\eeq
\section{Discussion}
Our dynamical results
%
can be compared to the analogous quenched values at similar lattice
spacing (corresponding to $\beta_{\rm quenched} = 6.0$):
\begin{eqnarray}
\left.
\begin{array}{c}
m^{\rm light}_{\overline{MS}}(2\, {\sf GeV}) = 5.5(5)\, {\sf MeV} \\
m^{\rm strange}_{\overline{MS}}(2\,{\sf GeV}) = 166(15)\, {\sf MeV}
\end{array}
\right\} & \mbox{{\rm quenched}} &\; .
\label{final2}
\end{eqnarray}
Errors due to the finite lattice spacing and the finite volume are not
included in (\ref{eqn_light_cond}, \ref{strange_final}) and (\ref{final2}).   
\par
Compared to the quenched result (at the same cutoff), which is in good
agreement with previous lattice calculations \cite{gupta}, we observe
a much smaller dynamical light quark mass, whereas the  
strange masses are compatible within errors. It is important to keep
in mind that the dynamical light and strange quark masses have been
obtained with two slightly different methods: the light quark is
obtained in two-flavour QCD whereas evaluating the strange quark mass
necessitates using valence quarks unequal to the sea quarks. This may
explain why the ratio of light to strange quark masses, so tightly
constrained by $\chi PT$, is not reproduced by the two dynamical
values of eqs.~\ref{eqn_light_cond} and \ref{strange_final}. 
\par
To illucidate this matter further, it is instructive to examine the
values of the light quark masses at fixed sea quark masses. 
The surprising result is that extrapolating the valence quark to 0 is not
sufficient to make the pseudoscalar masses $m_{PS}^2(sv)$ and
$m_{PS}^2(vv)$ vanish at fixed sea quark masses. Instead, we need to
tune the valence quark masses to negative values. The reason for this
can be traced back to the definition of the quark masses, which, for
Wilson-like theories, where chiral symmetry is broken, are definded
relative to the critical value of the hopping parameter
$\kappa_c$. In the dynamical theory, the critical point is
taken at vanishing sea quark, $\kappa_c = \kappa_c(\mqs = 0)$, $m_q =
{1 \over 2} \left( {1 \over \kappa} - {1 \over \kappa_c} \right)$,
where the mass of the pseudoscalar truely vanishes. 
If instead, we define a variable, sea quark mass dependent
 critical kappa at each sea quark value
we find much larger quark masses ($5.7(4)$, $5.6(3)$, $5.4(3)$ {\sf
  MeV}) which are similar to the quenched quark mass. The situation is
displayed schematically in figure \ref{geom}. $\Delta_1$ is the value
of the light quark mass we obtain in the full theory from $\mpsot$
whereas values such as $\Delta_2$ are obtained from fixed sea-quark
measurements (partially quenched). We see from figure \ref{geom} that the
light quark masses from finite sea-quark mass values do not
extrapolate to the value $\Delta_1$. Thus, to extract the light quark
mass, it it necessary to work with $\mpsot$; the data for $\mpsoth$
and $\mpstf$ can be used as a consistency check but are not
sufficient to obtain $\mql$. 
\par
To consolidate our results we are presently calculating the quark
masses using an alternative method in which the bare quark mass is
defined by means of the PCAC relation, 
\beq
m_q = { \partial_4 \left< A_4(x)P(0) \right> \over 2 \left< P(x) P(0)
\right> } \, , 
\eeq
where $A$ is the axial current and $P$ the pseudoscalar density. This
method may be less sensitive to renormalisation effects as the
previous one, thus giving an important cross-check. 
\begin{figure}[tb]
\epsfxsize=8cm
\epsfbox{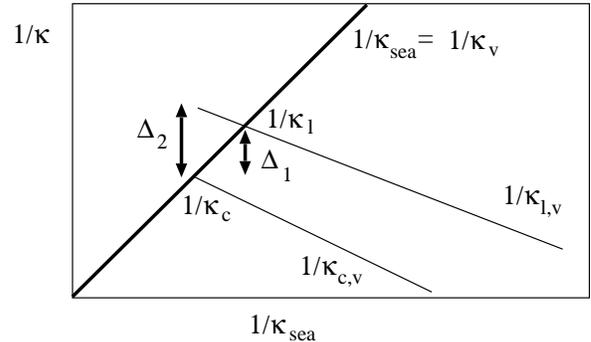}
\vspace{-6.5cm}
\caption{Schematic plot of light quark mass determinations (see
  text).\label{geom}}
\vspace{-0.5cm}
\end{figure}
%


\end{document}